\documentclass[fleqn,twoside]{article}

\usepackage{amssymb}
\usepackage{graphicx}

\usepackage{amsmath}

\begin{document}
\noindent
{\sf University of Shizuoka}

\vspace{2mm}

\begin{center}

%
{\Large\bf What Happens If an Unbroken Flavor Symmetry Exists?}
\vspace{2mm}

{\bf Yoshio Koide}

{\it Department of Physics, University of Shizuoka, 
52-1 Yada, Shizuoka 422-8526, Japan\\
E-mail address: koide@u-shizuoka-ken.ac.jp}

\end{center}

\vspace{3mm}
\begin{abstract}
Without  assuming any specific flavor symmetry and/or 
any specific mass matrix forms, it is demonstrated that 
if a flavor symmetry (a discrete symmetry, a U(1) symmetry, 
and so on) exists, we cannot obtain the CKM quark mixing matrix 
$V$ and the MNS lepton mixing matrix $U$ except for those 
between two families for the case with the completely undegenerated 
fermion masses, so that we can never give the observed CKM 
and MNS mixings. Only in the limit of $m_{\nu 1} =m_{\nu 2}$ 
($m_d=m_s$), we can obtain three family mixing with an
interesting constraint $U_{e3}=0$ ($V_{ub}=0$).
\end{abstract}



\vspace{5mm}


\noindent{\large\bf 1. Introduction} \ 

It is well known that the masses of the charged
fermions rapidly increase as $(u,d,e) \rightarrow
(c,s,\mu) \rightarrow (t,b,\tau)$.
It has been considered that the rapid increasing
of the mass spectra cannot be understood from
an idea of ``symmetry".
The horizontal degree of freedom has been called as
``generations".
In contrast to the idea of ``generations", there is
an idea of ``families" that the horizontal quantum
number states have basically the same opportunity.
It is after the democratic mass matrix model \cite{democ} 
was proposed that the idea of ``families" became 
one of the promising viewpoints for ``flavors".
Nowadays, a popular idea to understand the observed 
quark and lepton mass spectra and mixing matrices is 
to assume a flavor symmetry which puts constraints on
the Yukawa coupling constants.

In the present paper, we will point out that
if a flavor symmetry (a discrete symmetry, 
a U(1) symmetry, and so on) exists, we cannot obtain
the observed  Cabibbo-Kobayashi-Maskawa \cite{CKM} 
(CKM) quark mixing matrix $V_q$ and  
Maki-Nakagawa-Sakata \cite{MNS} (MNS) lepton mixing 
matrix $U_\ell$, even if we can obtain reasonable 
mass spectra under the symmetry. 
You may think that this conclusion is not so remarkable
and rather trivial,
because anyone thinks that the flavor symmetry
is badly broken.
However, most investigations on the broken flavor
symmetries are based on specific models, and
we are not clearly aware that what problem
happens if a flavor symmetry, in general, exists
until a low energy scale $\mu \sim 10^2$ GeV.
In the present paper,  
without assuming any explicit flavor symmetry and/or 
any explicit mass matrix forms, we will demonstrate 
how it is serious.  

Even when we consider a broken flavor symmetry, it is
important to consider the world in which the flavor
symmetry is unbroken.  In the present paper, we will
conclude that in such a world with an unbroken flavor
symmetry, the CKM and MNS mixing matrices cannot 
describe flavor mixings except for those between
two families when the fermion masses are completely
different from each other, and that only when
$m_{\nu 1}=m_{\nu 2}$ ($m_d=m_s$), the MNS matrix
$U_\ell$ (the CKM matrix $V_q$) can describe a three family
mixing with an interesting constraint $(U_\ell)_{e3}=0$
($(V_q)_{ub}=0$). 
This will suggests that our world with a broken flavor
symmetry should be derived from what unbroken world.

In the derivation of the conclusion, a requirement that
the SU(2)$_L$ symmetry must not be broken plays an
essential role.
Generally, the terminology ``symmetry" can have
a meaning only by defining the world in which
the symmetry is exactly unbroken.
In some of phenomenological mass matrix models, 
flavor symmetry breaking terms are brought into the theory
by hand, and it is not clear whether the terms can be
generated without breaking the SU(2)$_L$ or not.
In the present paper, we regard such a model with
an ad hoc flavor symmetry breaking as a model without a flavor
symmetry, and 
we will comment only on a model where 
the SU(2)$_L$ symmetry is exactly unbroken at the original 
Lagrangian, and the flavor symmetry breaking  mechanism does 
not spoil the SU(2)$_L$ invariance.

First, let us consider that the up- and down-quark fields
transform under a flavor symmetry as 
\begin{eqnarray}
    & u_L = U_{XL}^u u'_L, \ \ \ \ u_R = U_{XR}^u u'_R, 
\nonumber\\ 
    & d_L = U_{XL}^d d'_L,  \ \ \ \    d_R = U_{XR}^d d'_R. 
\end{eqnarray}
If the Lagrangian is invariant under the transformation (1),
the Yukawa coupling constants $Y_u$ and $Y_d$ must satisfy
the relations
\begin{equation}
(U_{XL}^u)^\dagger Y_u  U_{XR}^u = Y_u , \ \ \ 
(U_{XL}^d)^\dagger Y_d  U_{XR}^d = Y_d ,
\end{equation}
where $U_{XL}^u (U_{XL}^u)^\dagger = {\bf 1}$, and so on.
Since these transformations must not break SU(2)$_L$ symmetry,
we cannot consider a case with $U_{XL}^u \neq U_{XL}^d$.
We must rigorously take
\begin{equation}
U_{XL}^u = U_{XL}^d \equiv U_X .
\end{equation}
Therefore, the up- and down-quark mass matrices
$M_u=Y_u \langle H_u^0\rangle$ and 
$M_d=Y_d \langle H_d^0\rangle$ must satisfy the relations
\begin{equation}
U_X^\dagger M_u M_u^\dagger U_X = M_u M_u^\dagger, \ \ \ 
U_X^\dagger M_d M_d^\dagger U_X = M_d M_d^\dagger,
\end{equation}
independently of $U_{XR}^u$ and $U_{XR}^d$.

Similar situation is required in the lepton sectors.
Although, sometimes, in the basis where the charged
lepton mass matrix $M_e$ is diagonal (i.e.
$M_e = D_e \equiv {\rm diag}(m_e, m_\mu, m_\tau)$),
a ``symmetry" for the neutrino mass matrix $M_\nu$ is
investigated, such a prescription cannot be regarded as 
a field theoretical symmetry.
For example, when we assume a permutation symmetry
between neutrinos $\nu_{L2}$ and $\nu_{L3}$,
we can obtain a nearly bimaximal mixing \cite{23sym}.
However, the symmetry is applied only to neutrino sector $M_\nu$,
and not to the charged lepton sector $M_e=D_e$.
Therefore, we cannot regard this $2\leftrightarrow 3$ permutation
rule as a ``symmetry" in the field theoretical
meaning, because it is badly broken the SU(2)$_L$ symmetry.

In the lepton sectors, we must consider that
under the transformations
\begin{eqnarray}
    & \nu_L = U_{X} \nu'_L, \ \ \  \nu_R = U_{XR}^\nu \nu'_R, 
\nonumber\\ 
    & e_L = U_{X} e'_L,  \ \ \ \    e_R = U_{XR}^e e'_R, 
\end{eqnarray}
the Yukawa coupling constants which are defined by 
$\overline{e}_L Y_e e_R$, $\overline{\nu}_L Y_D^\nu e_R$, and
$\overline{\nu}_R^c Y_M^\nu \nu_R$  
($\nu_R^c \equiv C \overline{\nu}_R^T$) are invariant 
as follows
\begin{eqnarray}
       & U_{X}^\dagger Y_e U_{XR}^e = Y_e, \nonumber \\
       & U_{X}^\dagger Y_D^\nu U_{XR}^\nu = Y_D^\nu, \\
       & U_{XR}^T Y_M^\nu U_{XR} = Y_M^\nu.  \nonumber
\end{eqnarray}
In other words, the mass matrices $M_e M_e^\dagger$ and $M_\nu$
are invariant under the transformation $U_{X}$  as
\begin{eqnarray}
 & U_X^\dagger M_e M_e^\dagger U_X= M_e M_e^\dagger ,\\
 & U_X^\dagger M_\nu U_X^* = M_\nu ,
\end{eqnarray}
independently of the forms $U_{XR}^\nu$ and $U_{XR}^e$,
where we assumed the seesaw mechanism \cite{seesaw}
$M_\nu \propto Y_D^\nu (Y_M^\nu)^{-1} (Y_D^\nu)^T $.
(Even when we do not assume the seesaw mechanism, as long as 
the effective neutrino mass matrix is given by 
$\overline{\nu}_L M_\nu \nu_L^c$,
the mass matrix must obey the constraint (8).)

Note that the constraints (4) [and also (7) and (8)] do not always 
mean that the matrix forms $M_u M_u^\dagger$ and $M_d M_d^\dagger$ 
are identical each other.
Indeed, in the present paper, we consider a general case in which
the eigenvalues and mixing matrices between $M_u M_u^\dagger$ and 
$M_d M_d^\dagger$ are different from each other.
Nevertheless, the conditions (4) [and also (7) and (8)] will put 
very strong constraints on the CKM mixing matrix $V_q=(U_L^u)^\dagger U_L^d$ 
[and also the MNS mixing matrix $U_\ell=(U_L^e)^\dagger U_L^\nu$], 
where $U_L^f$ ($f=u,d,e,\nu$) are defined by
\begin{eqnarray}
&(U_L^f)^\dagger M_f M_f^\dagger U_L^f = D_f^2 \equiv
{\rm diag}( m_{f1}^2, m_{f2}^2,m_{f3}^2)  \ \ (f=u,d,e), \\
&(U_L^\nu)^\dagger M_\nu (U_L^\nu)^* = D_\nu \equiv
{\rm diag}( m_{\nu 1}, m_{\nu 2},m_{\nu 3}) .
\end{eqnarray}

The purpose of the present paper is 
to see whether it is possible or not to consider such the flavor 
symmetry without an SU(2)$_L$ symmetry breaking.
Of course, further conditions
\begin{equation}
(U_{XR}^f)^\dagger M_f^\dagger M_f U_{XR}^f = M_f^\dagger M_f
\ \ \ (f=u,d,e) ,
\end{equation}
will give more strict constraints on the mass matrices $M_f$.
However, even apart form such an additional constraint,
by using only the constraints (4), (7) and (8), we will
obtain a severe conclusion that such a symmetry cannot 
lead to the observed CKM mixing matrix $V_q$ and 
MNS mixing matrix $U_\ell$.

\vspace{2mm}

\noindent{\large\bf 2. Trouble in the CKM and MNS mixing matrices}\ 

First, we investigate relations in the quark sectors under 
the conditions (4).
Since we can rewrite the left hand of Eq.~(9) by using Eq.~(4) as
\begin{equation}
(U_L^f)^\dagger M_f M_f^\dagger U_L^f =
(U_L^f)^\dagger U_X^\dagger M_f M_f^\dagger U_X U_L^f =
(U_L^f)^\dagger U_X^\dagger U_L^f D_f^2 (U_L^f)^\dagger U_X U_L^f
,
\end{equation}
for $f=u,d$,
we obtain the relation 
\begin{equation}
(U_X^f)^\dagger D_f^2 U_X^f= D_f^2 , 
\end{equation}
where
\begin{equation}
U_X^f = (U_L^f)^\dagger U_X U_L^f .
\end{equation}
Therefore, the matrix $U_X^f$ which satisfies Eq.~(13) must 
be a diagonal matrix with a form
\begin{equation}
U_X^f = P_X^f \equiv {\rm diag}( e^{i \delta_1^f},
e^{i \delta_2^f}, e^{i \delta_3^f}) ,
\end{equation}
unless the masses are not degenerated.
Therefore, from (14), we obtain
\begin{equation}
U_X = U_L^u P_X^u (U_L^u)^\dagger = U_L^d P_X^d (U_L^d)^\dagger ,
\end{equation}
which leads to a constraint on the CKM matrix
$V_q \equiv (U_L^u)^\dagger U_L^d$:
\begin{equation}
P_X^u = V_q P_X^d (V_q)^\dagger .
\end{equation}
The constraint (17) (i.e. $P_X^u V_q = V_q P_X^d$) requires
\begin{equation}
(e^{i \delta_i^u } -e^{i \delta_j^d }) 
(V_q)_{ij} =0  \ \ (i,j=1,2,3).
\end{equation}
Only when $\delta_i^u = \delta_j^d$, we can
obtain $(V_q)_{ij} \neq 0$.
For the case $\delta_1^u =\delta_2^u =\delta_3^u \equiv \delta_u$
(also $\delta_1^d =\delta_2^d =\delta_3^d \equiv \delta_d$),
the matrix $P_X^u = {\bf 1} e^{i \delta_u}$ (and also 
$P_X^d = {\bf 1} e^{i \delta_d}$) leads to a trivial result
$U_X ={\bf 1} e^{i \delta_u}= {\bf 1} e^{i \delta_d}$), so
that we do not consider such a case.
Therefore, from the requirement (17), we cannot consider such a case
as all elements of $V_q$ are not zero.
For example, if we can take $(V_q)_{ii} \neq 0$ for $i=1,2,3$ by
taking $\delta_i^u =\delta_i^d \equiv \delta_i$,  since
we must choose, at least, one of $\delta_i$ differently from others,
we obtain  a mixing matrix  between only two families, e.g.
$(V_q)_{13}=(V_q)_{31}=(V_q)_{23}=(V_q)_{32}=0$ 
for a case of $\delta_1=\delta_2\neq \delta_3$:
\begin{equation}
V =\left(
\begin{array}{ccc}
\ast & \ast & 0 \\
\ast & \ast & 0 \\
0 & 0 & 1 
\end{array} \right) .
\end{equation}
Thus, for any choice of $\delta_i^u$ and $\delta_j^d$,
the condition (18) cannot lead to the observed CKM mixing matrix.

For the lepton sectors, the situation is the same.
{}From Eqs.~(8) and (10), we obtain the constraint
\begin{equation}
(U_X^\nu)^\dagger D_\nu (U_X^\nu)^* = D_\nu ,
\end{equation}
where
\begin{equation}
U_X^\nu = (U_L^\nu)^\dagger U_X U_L^\nu .
\end{equation}
Again, if we assume that the neutrino masses are not degenerated,
we obtain that the matrix $U_X^\nu$ must be diagonal, and
it is given by
\begin{equation}
U_X^\nu = P_X^\nu \equiv {\rm diag}( e^{i \delta_1^\nu},
e^{i \delta_2^\nu}, e^{i \delta_3^\nu}) ,
\end{equation}
because the constraint (20) leads to
\begin{equation}
\left(m_{\nu i} e^{-i \phi_{ij}} 
-m_{\nu j} e^{i\phi_{ij}}\right) |(U_X^\nu)_{ij}| = 0 ,
\end{equation}
where we have put $(U_X^\nu)_{ij}=  |(U_X^\nu)_{ij}| e^{i\phi_{ij}}$.
Here, differently from the matrix (15), 
the phases $\delta_i^\nu$ are
constrained as $\delta_i^\nu=0$ or $\delta_i^\nu= \pi$ 
($i=1,2,3$) from the condition (20).
{}From the relations (14) and (21), we obtain
\begin{equation}
U_X = U_L^e P_X^e (U_L^e)^\dagger = U_L^\nu P_X^\nu (U_L^\nu)^\dagger ,
\end{equation}
so that the MNS matrix $U_\ell = (U_L^e)^\dagger U_L^\nu$ must satisfy
the constraint
\begin{equation}
P_X^e = U_\ell P_X^\nu (U_\ell)^\dagger ,
\end{equation}
i.e.
\begin{equation}
(e^{i \delta_i^e } -e^{i \delta_j^\nu }) 
(U_\ell)_{ij} =0  \ \ (i,j=1,2,3).
\end{equation}
Again, only when $\delta_i^e = \delta_j^\nu$, we can
obtain $(U_\ell)_{ij} \neq 0$, and we cannot consider 
a case in which all elements of $U_\ell$ are not zero.
We only obtain  a mixing matrix  between two families.

Thus, the requirements (4) [and also (7) and (8)] lead to
a serious trouble in the CKM matrix  $V_q$ 
(the MNS matrix $U_\ell$),
even if we can suitably give the observed mass spectra. 
The similar conclusion has already been derived by Low and
Volkas \cite{Volkas} although they have demonstrated it by
using explicit mass matrix forms.

\vspace{2mm}
\noindent{\large\bf 3. Should we abandon any flavor symmetry?}\ 

In order to evade the conclusion (18) [and also
the conclusion (26)], we may consider a case with
$U_{XL}^u \neq U_{XL}^d$ [$U_{XL}^e \neq U_{XL}^\nu$].
However, such a transformation breaks SU(2)$_L$,
so that it is highly unrealistic.

If there is no symmetry breaking term in the
original Lagrangian, even if we take the
renormalization group equation (RGE) effects
into consideration, the SU(2)$_L$ is never broken,
and the relations (4), (7) and (8)
are still unchanged.

If we consider a U(1) charge model, we cannot
assign  different charges to $u_{Li}$ and $d_{Li}$
[and also to $\nu_{Li}$ and $e_{Li}$], so that we
must take the operator $U_X$ as
\begin{equation}
U_{XL}^u =U_{XL}^d \equiv U_X ={\rm diag}
(e^{i Q_1 \theta},e^{i Q_2 \theta},e^{i Q_3 \theta}) .
\end{equation}
In this case, since the Higgs scalars $H_u$ and $H_d$
can have different charges, the mass terms
$\overline{u}_L M_u u_R H_u$ and 
$\overline{d}_L M_d d_R H_d$ 
can have different phases for the transformation.
However, since the additional phases form Higgs sector
are common for all flavors, the conclusion (18) is 
essentially unchanged.

Related to an extended version of the U(1) charge model,
we know the Froggatt an Nielsen model \cite{Froggatt}.
The model can evade the present constraints (18) and (26).
In this model, each flavor state at a low energy 
scale has a different hierarchical structure, so that
the fermion flavors are ones which should be understood  
from the concept of ``generations" rather than 
from that of ``families".
The constraints in the present paper cannot be applied
to a model with  ``generation" structures, and
the  Froggatt and Nielsen model is indeed one of the most
promising models which can reasonably understand the generations.

Thus, it is one way to adopt a model with no flavor symmetry
in order to evade the present severe conclusions (18) and (26).
However, we know the fact (the degree of freedom
of ``rebasing") that 
we cannot physically distinguish two mass matrix sets 
$(M_u, M_d)$ and $(M'_u, M'_d)$, where $(M'_u, M'_d)$ 
is obtained from $(M_u, M_d)$ by a common flavor-basis 
rotation for the SU(2)$_L$ doublet fields. 
(The situation is the same in the lepton sector.)
Only when there is a flavor symmetry, the mass matrix forms
$(M_u, M_d)$ in a specific flavor basis have a meaning, 
because the operator of the flavor rotation does not 
commute with the flavor symmetry operator $U_X$.
Therefore,  the idea of a flavor symmetry is still attractive
to most mass-matrix-model-builders.

\vspace{2mm}
\noindent{\large\bf 4. Case of $m_{\nu 1} = m_{\nu 2}$}\ 

In order to seek for a clue to a possible
symmetry breaking, let us go on
a phenomenological study.

Since the observed neutrino data \cite{solar,kamland,atm,K2K} have 
shown $\Delta m^2_{solar} \ll \Delta m^2_{atm}$,
it is interesting to consider a limit of 
$m_{\nu 1}= m_{\nu 2}$.
In this case, the conclusion (22) [and (26)] is 
not correct any more, because the constraint (20) 
allows a case with $(U_X^\nu)_{12} \neq 0$ and 
$(U_X^\nu)_{21} \neq 0$:
\begin{equation}
U_X^\nu = \left(
\begin{array}{ccc}
c & -s & 0 \\
s & c & 0 \\
0 & 0 & 1
\end{array} \right) \ ,
\end{equation}
or 
\begin{equation}
U_X^\nu = \left(
\begin{array}{ccc}
-c & s & 0 \\
s & c & 0 \\
0 & 0 & 1
\end{array} \right) \ ,
\end{equation}
where $c=\cos\theta$ and $s=\sin\theta$.
(Again, each element must be real.)
Therefore, we must check the relation
\begin{equation}
P_X^e = U_\ell U_X^\nu (U_\ell)^\dagger ,
\end{equation}
with the forms of (28) and (29) of $U_X^\nu$, 
instead of (22).

Now, we explicitly calculate $U_\ell U_X^\nu (U_\ell)^\dagger$ 
by using a general form of $U_\ell$
\begin{equation}
U_\ell = V P_M ,
\end{equation}
where
\begin{equation}
V = \left(
\begin{array}{ccc}
c_{13} c_{12} & c_{13} s_{12} & s_{13} e^{-i \delta} \\
-c_{23} s_{12} -s_{23} c_{12} s_{13} e^{i\delta} &
c_{23} c_{12} -s_{23} s_{12} s_{13} e^{i\delta} &
s_{23} c_{13} \\
s_{23} s_{12} -c_{23} c_{12} s_{13} e^{i\delta} &
-s_{23} c_{12} -c_{23} s_{12} s_{13} e^{i\delta} &
c_{23} c_{13} 
\end{array} \right) ,
\end{equation}
and $P_M$ is a Majorana phase matrix
\begin{equation}
P_M = {\rm diag}(e^{i\alpha}, e^{i\beta}, e^{i\gamma}) .
\end{equation}

For the case with the form (28) of $U_X^\nu$, we obtain
\begin{eqnarray}
(U_\ell U_X^\nu U_\ell^\dagger)_{12}
&=&  -c_{13}\left\{ c_{23}\left(c_{12}^2 e^{-i\phi} 
+ s_{12}^2 e^{i\phi}\right) s \right. \nonumber \\
& &\left. + s_{13} s_{23} \left[ (e^{i\phi} -e^{-i\phi}) c_{12} s_{12} s
+c-1 \right] e^{-i\delta} \right\}
\end{eqnarray}
\begin{eqnarray}
(U_\ell U_X^\nu U_\ell^\dagger)_{13}
&=&  c_{13}\left\{ s_{23}\left(c_{12}^2 e^{-i\phi} 
+ s_{12}^2 e^{i\phi}\right) s \right. \nonumber \\
& &\left. - s_{13} c_{23} \left[ (e^{i\phi} -e^{-i\phi}) c_{12} s_{12} s
+c-1 \right] e^{-i\delta} \right\}
\end{eqnarray}
\begin{eqnarray}
(U_\ell U_X^\nu U_\ell^\dagger)_{23}
&=&  c_{23} s_{23} \left[
(e^{i\phi} -e^{-i\phi}) c_{12} s_{12} (1+s_{13}^2) s
+ c_{13}^2 (1-c) \right] \nonumber \\
& & - s_{13}s \left[ s_{23}^2 (c_{12}^2 e^{-i\phi} + s_{12}^2 e^{i\phi}) 
e^{i\delta}
+  c_{23}^2 (c_{12}^2 e^{i\phi} + s_{12}^2 e^{-i\phi})e^{-i \delta} 
\right] ,
\end{eqnarray}
where $\phi=\beta-\alpha$.
If $c_{13}\neq 0$, there is no solution which gives zeros for 
all the elements (34) -- (36), except for a trivial solution with
$c=1$ (i.e. $U_X={\bf 1}$).
If $c_{13}=0$, there is a solution for suitable choice of $\phi$
and $\delta$, and then, the matrix $V$ takes the form
\begin{equation}
V=\left(
\begin{array}{ccc}
0 & 0 & e^{-i \delta} \\
\ast & \ast & 0 \\
\ast & \ast & 0
\end{array} \right) .
\end{equation}
Of course, the form (37) is ruled out.
Thus, the case (28) cannot lead to any interesting form of $U_\ell$.

On the other hand, for the case (29), we obtain
\begin{eqnarray}
(U_\ell U_X^\nu U_\ell^\dagger)_{12}
&=&  c_{13}\left\{ c_{23} \left[ (c_{12}^2 e^{-i\phi} - s_{12}^2 e^{i\phi}) s 
+ 2 c_{12} s_{12} c  \right] 
\right. \nonumber \\
& &\left. + s_{13} s_{23} \left[ 1+(c_{12}^2 -s_{12}^2)c  
-(e^{i\phi} +e^{-i\phi}) c_{12} s_{12} s
 \right] e^{-i\delta} \right\}
\end{eqnarray}
\begin{eqnarray}
(U_\ell U_X^\nu U_\ell^\dagger)_{13}
&=& - c_{13}\left\{ s_{23} \left[ (c_{12}^2 e^{-i\phi} - s_{12}^2 e^{i\phi}) s 
+ 2 c_{12} s_{12} c  \right] \right. \nonumber \\
& &\left. - s_{13} c_{23} \left[ 1+(c_{12}^2 -s_{12}^2)c  
-(e^{i\phi} +e^{-i\phi}) c_{12} s_{12} s
 \right] e^{-i\delta} \right\}
\end{eqnarray}
\begin{eqnarray}
(U_\ell U_X^\nu U_\ell^\dagger)_{23}
&=&  c_{23} s_{23} \left[
(e^{i\phi} +e^{-i\phi}) c_{12} s_{12} (1+s_{13}^2) s
+ c_{13}^2 -(c_{12}^2 -s_{12}^2)(1+s_{13}^2) c \right] \nonumber \\
& & + s_{13} s_{23}^2 \left[(c_{12}^2 e^{-i\phi} - s_{12}^2 e^{i\phi}) s 
+ 2 c_{12} s_{12} c \right] e^{i\delta} \nonumber \\
& &  -  s_{13} c_{23}^2 \left[(c_{12}^2 e^{i\phi} - s_{12}^2 e^{-i\phi})s
+ 2 c_{12} s_{12} c \right] e^{-i \delta}  .
\end{eqnarray}
The case can lead to a non-trivial solution for $s_{13}=0$, 
$\phi=\beta-\alpha=0$ and
\begin{equation}
\cos(2\theta_{12} +\theta) = 1 ,
\end{equation}
i.e.
\begin{equation}
U_\ell = \left(
\begin{array}{ccc}
c_{12} & s_{12} & 0 \\
-c_{23} s_{12} & c_{23} c_{12} & s_{23} \\
s_{23} s_{12} & -s_{23} c_{12} & c_{23} 
\end{array} \right) P_M .
\end{equation}
It should be noted that in the limit of $m_{\nu 1}=
m_{\nu 2}$, the Majorana phases in $P_M$ must be
$\alpha=\beta$.

The similar result $(U_\ell)_{13}=0$  has also been derived 
by Low and Volkas \cite{Volkas} although their interest was in
the ``trimaximal mixing" and they have assumed a specific
flavor symmetry.
In the present general study, we can obtain $s_{13}=0$,
but $s_{12}$ and $s_{23}$ are still free.
The result (42) is a conclusion which is derived 
model-independently.

Note that the case (29) satisfies $(U_X^\nu)^2 = {\bf 1}$,
so that the flavor transformation $U_X$ also satisfies
\begin{equation}
(U_X)^2 = {\bf 1} .
\end{equation}
This suggests  that an approximate flavor symmetry in the lepton sectors 
is a discrete symmetry Z$_2$.

Inversely, for the neutrino mass spectra with 
$m_{\nu 1} \neq m_{\nu 2}$, if we take the operator 
$U_X= U_L^\nu U_X^\nu (U_L^\nu)^\dagger$ with the form
(29) of $U_X^\nu$, we obtain
\begin{equation}
(U_X^\nu)^\dagger D_\nu (U_X^\nu)^* = D_\nu
+(m_{\nu 2} -m_{\nu 1}) s \left(
\begin{array}{ccc}
s & c & 0 \\
c & -s & 0 \\
0 & 0 & 0
\end{array} \right) ,
\end{equation}
which leads to
\begin{equation}
U_X^\dagger M_\nu U_X^* = M_\nu + (m_{\nu 2} -m_{\nu 1}) s B ,
\end{equation}
where the symmetry breaking term $B$ is given by
\begin{equation}
B= U_L^\nu \left(
\begin{array}{ccc}
s & c & 0 \\
c & -s & 0 \\
0 & 0 & 0
\end{array} \right) (U_L^\nu)^T .
\end{equation}
The matrix $B$ is rewritten as
\begin{equation}
B= U_L^e \left(
\begin{array}{ccc}
0 & c_{23} & -s_{23} \\
c_{23} & 0 & 0 \\
-s_{23} & 0 & 0
\end{array} \right) (U_L^e)^\dagger ,
\end{equation}
by using the relation $U_X = U_L^e P_X^e (U_X^e)^\dagger$
and the constraint (41).
Of course, the result (45) shows that in the limit of
$m_{\nu 1} = m_{\nu 2}$ and/or $s=0$, the operation
$U_X$ becomes that of the exact symmetry.
The forms (46) and (47) of the symmetry breaking term
will give a clue to a possible form of the flavor
symmetry breaking.
However, in order to fix the values of $s_{23}$ and
$s_{12}$ (or $s$), we must put a further assumption.
In the present paper, we do not give such a 
speculation any more.

If we apply the similar discussion to the quark sector
in the limit of $m_d=m_s$, we can obtain $|V_{ub}|=0$.
This may be taken as the reason of $|V_{ub}|^2 \ll 
|V_{cb}|^2, |V_{us}|^2$.

\vspace{2mm}
\noindent{\large\bf 5. Concluding remarks}\ 

In conclusion, we have noticed that when we assume
a flavor symmetry, we must use the same operation $U_X$
simultaneously for the up-quarks $u_{Li}$ and down-quarks $d_{Li}$
(and also for the charged leptons $e_{Li}$ and
neutrinos $\nu_{Li}$), and we have demonstrated that
the existence of such an operation $U_X$ without an SU(2)$_L$
breaking leads to unwelcome forms of the CKM mixing
matrix $V_q=(U_L^u)^\dagger U_L^d$ and the MNS mixing matrix 
$U_\ell (U_L^e)^\dagger U_L^\nu$,
even if we can obtain reasonable mass spectra:
in the limit of an unbroken flavor symmetry, the CKM and MNS 
mixing matrices cannot describe flavor mixings except for only 
those between two families when the fermion masses are completely
different from each other, and that only when
$m_{\nu 1}=m_{\nu 2}$ ($m_d=m_s$), the MNS matrix
$U_\ell$ (the CKM matrix $V_q$) can describe a three family
mixing with an interesting constraint $(U_\ell)_{e3}=0$
($(V_q)_{ub}=0$). 

If we want to investigate the ``generation" problem from the
standpoint of flavor symmetry, our results (18) and (26)
demands that the flavor symmetry should be completely
broken at a high energy scale $M_X$, so that we cannot 
have any flavor symmetry below $\mu=M_X$.
We have to seek for a flavor symmetry breaking mechanism under 
the condition that the original Lagrangian (including the
symmetry breaking mechanism) is exactly invariant
under the SU(2)$_L$.

For example, let us consider a two Higgs doublet model,
or a $\overline{5}_L \leftrightarrow \overline{5}'_L$ model
\cite{55mix}. 
In such a model, the effective Yukawa coupling constants
$Y^f$ below $\mu=M_X$ are given by a linear combination 
of two Yukawa coupling constants with different textures
$Y^f_A$ and $Y^f_B$,
\begin{equation}
Y^f = c_A^f Y_A^f +c_B^f Y_B^f ,
\end{equation}
so that $Y^f (Y^f)^\dagger$ do not satisfy the flavor
symmetry condition
\begin{equation}
U_X^\dagger Y^f (Y^f)^\dagger U_X = Y^f (Y^f)^\dagger ,
\end{equation}
although $Y^f_A (Y^f_A)^\dagger$ and $Y^f_B (Y^f_B)^\dagger$
must satisfy the conditions
\begin{equation}
U_X^\dagger Y^f_A (Y^f_A)^\dagger U_X = Y^f_A (Y^f_A)^\dagger ,
\ \ \ 
U_X^\dagger Y^f_B (Y^f_B)^\dagger U_X = Y^f_B (Y^f_B)^\dagger ,
\end{equation}
respectively, even if at $\mu< M_X$.
In other words, there is no operator $U_X$ which satisfies the
condition (49).
Thus, we can break the flavor symmetry without the SU(2)$_L$
symmetry.

However, we should note that the matrices $Y_A^f$ and $Y_B^f$
have to satisfy the conditions (50).
As an example, let see a two Higgs doublet model with Z$_3$
and S$_2$ symmetries \cite{univtex}. 
In the model, we assume that under a discrete symmetry Z$_3$, 
the quark and lepton fields $\psi_L$,
which belong to $10_L$, $\overline{5}_L$ and $1_L$ of SU(5) 
($1_L={\nu}_R^c$),
are transformed as
\begin{equation}
\psi_{1L} \rightarrow \psi_{1L} , \ \
\psi_{2L} \rightarrow \omega \psi_{2L} , \ \
\psi_{3L} \rightarrow \omega \psi_{3L}, 
\end{equation}
where $\omega=e^{2i\pi/3}$.
[Although we use a terminology of SU(5), at present,
we do not consider the SU(5) grand unification.]
Then, the bilinear terms $\overline{q}_{Li} u_{Rj}$, 
$\overline{q}_{Li} d_{Rj}$, $\overline{\ell}_{Li} \nu_{Rj}$, 
$\overline{\ell}_{Li} e_{Rj}$ and $\overline{\nu}_{Ri}^c \nu_{Rj}$
[$\nu_R^c =(\nu_R)^c =C \overline{\nu_R}^T$ and
$\overline{\nu}_R^c =\overline{(\nu_R^c)}$] 
are transformed as follows:
\begin{equation}
\left( 
\begin{array}{ccc}
1 & \omega^2 & \omega^2 \\
\omega^2 & \omega & \omega \\
\omega^2 & \omega & \omega \\
\end{array} \right) \ .
\end{equation} 
Therefore, if we assume two SU(2) doublet Higgs scalars $H_A$ and $H_B$, 
which are transformed as
\begin{equation}
H_A \rightarrow \omega H_A , \ \ \ H_B \rightarrow \omega^2 H_B , 
\end{equation}
we obtain Yukawa coupling constants with the following textures:
\begin{equation}
Y_A^f = a_f \left(
\begin{array}{ccc}
0 & 1 & 1 \\
1 & 0 & 0 \\
1 & 0 & 0
\end{array} \right) , \ \ \ 
Y_B^f = b_f
\left(
\begin{array}{ccc}
0 & 0 & 0 \\
0 & 1 & x_f \\
0 & x_f & 1
\end{array} \right) ,
\end{equation}
where in addition to the Z$_3$ symmetry, we have
assumed a flavor $2\leftrightarrow 3$ symmetry 
(S$_2$ symmetry).
At a high energy scale $\mu=M_X$, the Z$_3$ symmetry is
broken, and the mixing between $H_A$ and $H_B$ takes place.
We assume that the one component of the linear combinations
of $H_A$ and $H_B$ plays a role of the conventional Higgs 
scalar, so that we obtain the following universal texture of
quark and lepton mass matrices $M_f$:
\begin{equation}
M_f = P_f  \widehat{M}_f  P_f  \ ,
\end{equation}
where $P_f$ is a phase matrix defined by
\begin{equation}
P_f = {\rm diag} (e^{i \delta^{f}_1},\ e^{i \delta^{f}_2}, \
e^{i \delta^{f}_3}) \ ,
\end{equation}
and $\widehat{M}_f$ is a real matrix with a form
\begin{equation}
\widehat{M}_f =
\left(
\begin{array}{ccc}
0 & a_f & a_f \\
a_f & b_f & b_f x_f \\
a_f & b_f x_f & b_f
\end{array} \right)   \ ,
\end{equation}
[$a_f$ and $b_f$ have been redefined by including the mixing
coefficients, differently from those in the expressions (54)].
Here, note that the parameters $\delta_i^f$ are phenomenological
ones.
If we still require S$_2$ symmetry, we obtain $\delta_2^f =
\delta_3^f$.
If we assume an ``extended $2\leftrightarrow 3$ symmetry" 
with a phase conversion \cite{univtex_YK}, we can obtain  
$\delta_2^f \neq \delta_3^f$, but the SU(2)$_L$ symmetry still 
requires $\delta_3 -\delta_2 \equiv (\delta_3^u -\delta_3^d)
-(\delta_2^u -\delta_2^d) =0$.
On the other, in this model, the nonvanishing value of
$\delta_3-\delta_2$ is essential to give a nonvanishing
value of $|V_{cb}|$, because it is given by
$|V_{cb}| \simeq \sin(\delta_3-\delta_2)/2$ in the model.
Thus, in this model, we cannot obtain reasonable CKM and
MNS mixing matrices without breaking the SU(2)$_L$
symmetry (i.e. $\delta_3 -\delta_2 \neq 0$).
For currently proposed phenomenological mass matrix
models with a flavor symmetry breaking,
it is important to check whether the mass matrix
forms with a broken flavor symmetry are still invariant
or not under the SU(2)$_L$ symmetry.

Finally, we would like to comment on the results
(42) and (43) in the case of $m_{\nu 1} = m_{\nu 2}$.
This suggests a possibility that we can reasonable 
understand the observed smallness of $|(U_\ell)_{13}|$ 
\cite{CHOOZ} and $|(V_q)_{ub}|$ \cite{PDG02} if we consider 
a model with a flavor symmetry of Z$_2$ type, 
$(U_X)^2 ={\bf 1}$, and with 
$m_{\nu 1}= m_{\nu 2}$ and $m_{d1}=m_{d2}$ at
$\mu > M_X$. 
This will give a promising clue to possible features of
the unbroken flavor symmetry.

\vspace{7mm}

\centerline{\large\bf Acknowledgments}

The author would like to thank J.~Sato for helpful
conversations.
This work was supported by the Grant-in-Aid for
Scientific Research, the Ministry of Education,
Science and Culture, Japan (Grant Number 15540283).


\end{document}